# A numerical investigation on the influence of liquid properties and interfacial heat transfer during microdroplet deposition onto a glass substrate

**Rajneesh Bhardwaj[1], Jon P. Longtin[2] and Daniel Attinger[1, *]**

[1]Laboratory for Microscale Transport Phenomena,

Department of Mechanical Engineering,

Columbia University, New York, NY 10025

[2]Thermal-Laser Laboratory

Department of Mechanical Engineering,

State University of New York at Stony Brook, Stony Brook, NY 11794

*Corresponding author. Tel: +1-212-854-2841; fax: +1-212-854-3304;

E-mail address: da2203@columbia.edu (D. Attinger)

**Abstract**

This work investigates the impingement of a liquid microdroplet onto a glass substrate at different temperatures. A finite-element model is applied to simulate the transient fluid dynamics and heat transfer during the process. Results for impingement under both isothermal and non-isothermal conditions are presented for four liquids: isopropanol, water, dielectric fluid (FC-72) and eutectic tin-lead solder (63Sn-37Pb). The objective of the work is to select liquids for a combined numerical and experimental study involving a high resolution, laser-based interfacial temperature measurement to measure interfacial heat transfer during microdroplet deposition. Applications include spray cooling, micro-manufacturing and coating processes, and electronics packaging. The initial droplet diameter and impact velocity are 80 μm and 5 m/s, respectively. For isothermal impact, our simulations with water and isopropanol show very good agreement with experiments. The magnitude and rates of spreading for all four liquids are shown and compared. For non-isothermal impacts, the transient drop and substrate temperatures are expressed in a non-dimensional way. The influence of imperfect thermal contact at the interface between the drop and the substrate is assessed for a realistic range of interfacial Biot numbers. We discuss the coupled influence of interfacial Biot numbers and hydrodynamics on the initiation of phase change.

## Nomenclature

| | |
|---|---|
| $Bi$ | Biot number $[h_c d_0 k_l^{-1}]$ |
| $c$ | speed of sound [m s$^{-1}$] |
| $c_p$ | specific heat [J kg$^{-1}$K$^{-1}$] |
| $C$ | dimensionless heat capacity $[\rho c_p / \rho_l c_{p,\,l}]$ |
| $d$ | splat diameter [m] |
| $e$ | distance of center of circular measurement spot from origin on r-axis [m] |
| $Fr$ | Froude number $[v_0^2 d_0^{-1} g^{-1}]$ |
| $g$ | gravitational acceleration [9.81 m s$^{-2}$] |
| $h$ | interfacial heat transfer coefficient [Wm$^{-2}$K$^{-1}$] |
| $H$ | mean surface curvature [m$^{-1}$] |
| $\overline{H}$ | dimensionless mean surface curvature $[Hd_0]$ |
| $k$ | thermal conductivity [Wm$^{-1}$K$^{-1}$] |
| $K$ | dimensionless thermal conductivity $[k k_l^{-1}]$ |
| $M$ | Mach number $[v_0 c^{-1}]$ |
| $n$ | number of grid points inside circular measurement spot |
| $p$ | pressure [Pa] |
| $P$ | dimensionless pressure $[p v_0^{-2} \rho_l^{-1}]$ |
| $Pr$ | Prandtl number $[\mu c_{p,\,l} k_l^{-1}]$ |
| $q$ | heat flux at the splat/substrate interface [W m$^{-2}$] |
| $r$ | radial coordinate [m] |
| $R$ | dimensionless radial coordinate $[r d_0^{-1}]$ |
| $Re$ | Reynolds number $[\rho v_0 d_0 \mu^{-1}]$ |
| s | radius of spot [m] |
| $t$ | time [s] |
| $T$ | temperature [K] |
| $u$ | radial velocity [m $s^{-1}$] |
| $U$ | dimensionless radial velocity [$uv_0^{-1}$] |
| $v$ | axial velocity [m $s^{-1}$] |
| $V$ | dimensionless axial velocity [$vv_0^{-1}$] |
| $We$ | Weber number [$\rho v_0^2 d_0 \gamma^{-1}$] |
| $z$ | axial coordinate [m] |
| $Z$ | dimensionless axial coordinate [$zd_0^{-1}$] |

**Greek symbols**

| | |
|---|---|
| $\alpha$ | thermal diffusivity [$m^2s^{-1}$] |
| $\beta$ | spread factor [$d_{max}d_0^{-1}$] |
| $\delta$t | temporal resolution available by experimental setup |
| $\gamma$ | surface energy [$Jm^{-2}$] |
| $\phi$ | contact angle |
| $\mu$ | dynamic viscosity [Pa s] |
| $\theta$ | dimensionless temperature [$\{T - \min(T_{1,0}, T_{2,0})\}(\|T_{1,0}-T_{2,0}\|)^{-1}$] |
| $\rho$ | density [kg $m^{-3}$] |
| $\sigma$ | stress [Pa] |
| $\tau$ | dimensionless time [$tv_0d_0^{-1}$] |

**Subscripts**

| | |
|---|---|
| 0 | initial |
| 1 | drop/splat |
| 2 | substrate |
| avg | average value |
| c | contact, interface |
| i | initial |
| int | linearly interpolated value |
| *l* | liquid |
| max | maximum value |
| r | radial direction |
| z | axial direction |

# 1 Introduction

The fluid dynamics and heat transfer associated with microdroplet impingement onto a substrate are of considerable interest to micro-manufacturing, spray cooling, spray coating, and inkjet-printing [1-3]. A variety of fluids are used in such processes, including fuels in combustion, water and dielectric fluids for cooling, and metal droplets for rapid prototyping and electronic interconnects [4, 5].

In this work a numerical investigation of a liquid microdroplet impacting on a horizontal substrate at different temperatures is presented (Figure 1). The initial droplet diameter and impact velocity are 80 μm and 5 m/s respectively and gravity is negligible. The associated transport phenomena are extremely complex. For instance, this problem involves fluid dynamics with large deformations of the droplet free surface. The simultaneous, transient heat transfer process involves convection in the droplet coupled with conduction in the substrate. Both the thermal field inside the droplet and the onset of phase change, if any, depend on the interfacial heat transfer coefficient between the droplet and substrate, which expresses the imperfect thermal contact at the interface. Our study focuses on drops of eutectic tin-lead solder (63Sn-37Pb, referred as solder, hereafter), water, isopropanol and FC-72 fluorocarbon, a dielectric fluid used for electronics cooling. A primary objective of this work is to evaluate these liquids as potential candidates for a companion experimental study, currently being developed, to measure interfacial heat transfer coefficients and temperature history and compare the results to numerical simulations.

Traditionally, numerical models targeting similar problems use strong simplifications for the sake of numerical tractability [3]. For instance Harlow and Shannon [6] neglected both viscous and surface tension effects in their modeling of a liquid droplet impacting on a flat plate.

Tsurutani *et al*. [7] used the simplified marker and cell method (SMAC) and employed a fixed grid with relatively low resolution. Increasing computing capacities have recently led to very convincing simulations of the impact of millimeter-size drops with the Volume-Of-Fluid method [8], however the ability of this technique to address micrometer-size droplet cases, where free surface effects are more important, is not assessed yet. Gao and Sonin [9] developed a powerful theoretical analysis in which order-of-magnitude approximations were made to characterize the associated time scales, such as the times required to remove the initial superheat, remove the latent heat during freezing, and subsequently cool the deposit to the ambient temperature. Two effects that have been shown to be significant in other studies were neglected in this formulation: *convection effects* within the droplet and *thermal contact resistance* at the splat-substrate interface [10, 11].

Zhao *et al*. [12] modeled the cooling of a liquid microdroplet, accounting for fluid dynamics phenomena and assuming perfect interfacial thermal contact. This group used a Lagrangian formulation, extending the fluid dynamics model of Fukai *et al*. [13] to account for the heat transfer process in the droplet and substrate. Wadvogel *et al*. [14, 15] extended this modeling to account for solidification and imperfect interfacial thermal contact. This modeling is used in this article, with the incorporation of a more stable and versatile mesh generation scheme *Mesh2d* [16], and the ability to modify the interfacial heat transfer coefficient with respect to time and space.

Several studies have specifically investigated the role and importance of imperfect thermal contact between the substrate and the drop. This imperfect thermal contact is a critical parameter in the heat transfer process. Liu *et al*. [17] suggest that when a liquid spreads over a solid surface, perfect thermal contact cannot be achieved between the liquid and solid surface because of the substrate surface roughness, surface tension, surface impurities, and gas entrapment. It is believed that heat transfer through the actual (imperfect) contact area occurs by conduction and, to some degree, radiation across the gas-filled gaps [3]. For molten lead droplets, imperfect thermal contact was experimentally observed by Bennett and Poulikakos [11]. Pasandideh-Fard *et al.* [18] and Xiong *et al.* [19] performed a numerical study on the sensitivity to contact resistance on the final diameter, overall shape and height of a solidified solder droplet. Their model predicted variations in solders bump height up to 20% due to variations of thermal contact resistance. Recently, Attinger and Poulikakos [20] compared experimental and numerical transient oscillations for a solidifying solder drop and were able to estimate the value of the interfacial heat transfer coefficient for a specific case. Although the investigations above have shown the importance and effects of the interfacial heat transfer coefficient, there is still a lack of modeling and predictive tools to determine *a priori* the interfacial heat transfer coefficient.

This study is aimed at selecting liquids and temperature for a combined theoretical and experimental investigation of microdroplet fluid dynamics and heat transfer on a cooled surface. The laser-based technique developed by Chen et al. [21] will be modified and used to measure the interfacial temperature with microsecond and micrometer resolution under a spreading droplet. Matching the measured and calculated temperature values at the interface will allow the determination of the transient and local behavior of the heat transfer coefficient, which is a necessary step in developing predictive models for interfacial heat transfer. In this article, we discuss the effect of interfacial heat transfer on the heat transfer process during the impact of solder, water, isopropanol and FC-72 (dielectric fluid) droplets on a glass substrate.

## 2 Numerical Model

The mathematical model is based on the Navier-Stokes and energy equations [14] applied to an axisymmetric geometry. All equations are expressed in a Lagrangian framework, which provides accurate modeling of the large deformations of the free surface and the associated Laplace stresses [13].

### 2.1 Fluid dynamics

The flow inside the droplet is laminar and all thermophysical properties are assumed to be constant with respect to temperature. The radial and axial components of the momentum equation are considered along with the continuity equation. An artificial compressibility method is employed to transform the continuity equation into a pressure evolution equation. This method assumes a fluid flow that is slightly compressible, whereby the speed of sound is large, but not infinite. A Mach number of 0.001 is used for all simulations in this work. The derivation of the boundary condition at the free surface consider forces due to pressure, viscous stresses and surface tension [13]. The traditional no-slip boundary condition fails in the vicinity of the contact line because its application results in an infinite stress in the region. To circumvent this problem, a scheme proposed by Bach and Hassager [22] is utilized, which applies a net interfacial force given by the equilibrium surface tension coefficient of the joining phases. The wetting force at the dynamic contact line between the liquid droplet and the substrate is neglected throughout the analysis. The dimensional form of the fluid dynamics equations can be found in [23], with the expression of the stress tensor. The dimensionless equations for fluid dynamics are [14]:

*Mass conservation:*

$$\frac{\partial P}{\partial \tau}+\frac{1}{M^2}\left(\frac{1}{R}\frac{\partial}{\partial R}\left(RU+\frac{\partial V}{\partial Z}\right)\right)=0 \qquad (1)$$

where $P$, $\tau$, $R$, $Z$, $U$, $V$ are dimensionless pressure, time, radial distance, axial distance, radial velocity and axial velocity, respectively. $M$ denotes Mach number.

*Momentum conservation in radial direction:*

$$\frac{\partial U}{\partial \tau} - \frac{1}{R}\frac{\partial}{\partial R}\left(R\overline{\sigma}_{RR}\right) - \frac{\partial \overline{\sigma}_{RZ}}{\partial Z} + \frac{1}{R}\overline{\sigma}_{\theta\theta} = 0 \tag{2}$$

where $\overline{\sigma}_{RR}$ and $\overline{\sigma}_{ZZ}$ are dimensionless stress tensor terms.

*Momentum conservation in axial direction:*

$$\frac{\partial V}{\partial \tau} - \frac{1}{R}\frac{\partial}{\partial R}\left(R\overline{\sigma}_{RR}\right) - \frac{\partial \overline{\sigma}_{RZ}}{\partial Z} + \frac{1}{Fr} = 0 \tag{3}$$

In the above equation, $Fr$ denotes Froude number.

## 2.2 Heat transfer

The energy equation is solved in both the droplet and the substrate, according to the formulation in [14]. Convection and radiation heat transfer from all exposed surfaces is neglected. The dimensionless energy conservation equation for droplet and substrate is given by ($i$ = 1 for droplet and $i$ = 2 for substrate):

$$C_i \frac{\partial \theta_i}{\partial \tau} - \frac{1}{\Pr \operatorname{Re}}\left[\frac{1}{R}\frac{\partial}{\partial R}\left(K_i R \frac{\partial \theta_i}{\partial R}\right) + \frac{\partial}{\partial Z}\left(K_i \frac{\partial \theta_i}{\partial Z}\right)\right] = 0 \tag{4}$$

where $C_i$ and $K_i$ is the dimensionless heat capacity and thermal conductivity respectively. $Pr$ and $Re$ denotes Prandtl and Reynolds number respectively. $\theta_i$ is the dimensionless temperature, and is defined as:

$$\theta_i = \frac{T_i - \min(T_{1,0}, T_{2,0})}{\left|T_{1,0} - T_{2,0}\right|} \tag{5}$$

where $T_{1,0}$ and $T_{2,0}$ are the initial dimensional temperature of drop and substrate respectively.

### 2.3 Thermal contact resistance

Thermal contact resistance between droplet and substrate is modeled by a thin layer of arbitrary thickness $\delta$, with zero heat capacity and adjustable thermal conductivity $k_i$ [19]. The interfacial heat transfer coefficient can therefore be defined as $h_c = k_i/\delta$. This approach is fully compatible with that of Wang and Matthys [24]. The thermal contact resistance can be non-dimensionalized with the Biot number as [19]:

$$Bi = \frac{h_c d_0}{k_l} \tag{6}$$

where, $d_0$ is the initial diameter of the droplet.

### 2.4 Initial and Boundary conditions

The initial conditions are as follows:

$$U = 0; V = -1; P = \frac{4}{We} \tag{7}$$

$$\theta_1(R,Z,0) = 1; \theta_2(R,Z,0) = 0 \quad \text{for solder} \tag{8}$$

$$\theta_1(R,Z,0) = 0; \theta_2(R,Z,0) = 1 \quad \text{for water, isopropanol and FC-72} \tag{9}$$

The last two initial conditions show that the solder drop is cooled upon contact with the substrate, while water, isopropanol and FC-72 drops are heated.

The boundary conditions are as follows:

$$U = 0; \frac{\partial V}{\partial R} = 0 \quad \text{at } R = 0 \tag{10}$$

$$U = V = 0 \quad \text{at } Z = 0 \tag{11}$$

$$\overline{\sigma_{RR}}n_R + \overline{\sigma_{RZ}}n_Z = -2\frac{\overline{H}}{We}n_R \qquad \text{at droplet free surface} \tag{12}$$

$$\overline{\sigma_{ZR}}n_R + \overline{\sigma_{ZZ}}n_Z = -2\frac{\overline{H}}{We}n_Z \qquad \text{at droplet free surface} \tag{13}$$

The above two boundary conditions are the balance of forces due to pressure, viscous stresses and surface tension at droplet free surface.

$$\frac{\partial \theta_i}{\partial R}n_r + \frac{\partial \theta_i}{\partial Z}n_z = 0 \qquad \text{at droplet free surface and the substrate boundary surface} \tag{14}$$

## 3 Numerical Scheme

The computational domain is discretized as a mesh of triangular elements and the numerical model is solved using a Galerkin finite element method. Linear shape functions are used for velocity and pressure. An implicit method is utilized for the integration of fluid dynamics equations in time, while a Crank-Nicholson scheme is used for the energy equation. Details of the algorithm are given in [15]. The present model uses a more robust and freely available mesh generator *Mesh2D* [16]. It has been found that *Mesh2D* is better than the advancing front method [25] in terms of the time taken to generate mesh, the allowable aspect ratio of the elements, and the number of elements generated. A comparison of meshes generated by two methods is shown in Figure 2.

In the present work, the grid and time step independence are examined in terms of the position of the contact point between the *z*-axis and the free surface of the splat ($Z_c$). This study is carried out for a 50 μm solder droplet impacting a flat surface at 2.0 m/s under isothermal conditions. This corresponds to *Re* = 314 and *We* = 4.76. The grid independence is considered for four increasing number of nodes in the droplet: 199, 521, 705 and 873, with a time step of $5\times10^{-4}$ in each case. The time step independence is considered for time steps of 0.5, 1, 2, and 3 $\times10^{-3}$ for 705 nodes in each case, with the results shown in Figure 3. As it can be seen, a time step of $5\times10^{-4}$ and a spatial discretization of 700 nodes in the droplet are sufficient for the simulations. Each simulation requires approximately 6 CPU hours on a 2.4 GHz Intel-Xeon machine with 1 GB of RAM.

### 3.1 Thermophysical properties and dimensionless numbers

The thermophysical properties and dimensionless numbers used for simulations are given in Table 1 and Table 2 respectively.

## 4 Results and Discussions

Results are presented for solder, water, isopropanol and FC-72 droplets with diameter $d_o$ = 80 μm; velocity $v_o$ = 5 m/s and values of Bi of 1, 10 or 100. This choice of Biot numbers represents a realistic range of values used in previous work [19]. The initial temperatures are dimensionless, which means that a single simulation result describes any non-isothermal impact. In case of solder the droplet is cooled by the substrate, so the initial dimensionless temperatures for drop and substrate are 1 and 0 respectively. In the cases where water, isopropanol and FC-72 droplets are heated by the substrate, these corresponding values are 0 and 1.

### 4.1 Fluid dynamics

Figure 4 shows the spreading of a solder microdroplet with successive representations of the droplet shape, temperature isotherms, and streamlines. During the initial spreading stage ($t$ < 12 μs), the deformation of the drop is mostly influenced by inertial forces. However, in the later stages of spreading ($t$ > 25 μs), inertial forces decrease and surface tension forces dominate. This competition between inertial and surface tension forces induces the peripheral ring visible for $t$ =

20 μs, as well as a strong recoiling which results in the splashing of the solder drop. Also a vortex forms in the drop during recoiling (Figure 4).

### 4.1.1 Comparison of spreading in all four liquids

The temporal evolution of the spread factor $\beta$ (ratio of maximum splat diameter to initial droplet diameter) for all four liquids is plotted in Figure 5. The least spreading is observed with solder, which due to its small Weber number (Table 2) and the high viscous dissipation. The maximum spreading occurs with FC-72 because of its large Weber number. In general, a larger Weber number results in more substantial droplet spreading.

### 4.1.2 Comparison with previous results

Recently our numerical code was validated with experimental results for solder [20]. In the present work, numerical values of the maximum spread factor for water and isopropanol are compared with visualization results [26], and also with analytical expressions available in the literature [18]. In Table 3, we use the same parameters as in the visualization study [26]: for isopropanol the parameters are $d_0$ = 87 μm and $v_0$ = 9.28 m/s (Re = 259, We = 277). For water, the parameters are: $d_0$ = 83 μm and $v_0$ = 8.19 m/s (Re = 696, We =77). The analytical estimate for the maximum value of the spread factor ($\beta_{max}$) in [18] assumes that the surface energy at the maximum spreading equals the kinetic and surface energy before impact, less the viscous dissipation during impact:

$$\beta_{\max} = \frac{d_{\max}}{d_0} = \sqrt{\frac{We + 12}{3(1 - \cos\phi\ ) + 4\dfrac{We}{\sqrt{\mathrm{Re}}}}} \tag{15}$$

where, $\phi$ is contact angle.

Table 3 shows very good agreement between numerical, experimental and analytical results, except for the maximum spread factor of isopropanol. This may be explained by the fact that the lower Reynolds number related to the isopropanol impact does not fully match the assumption in Equation 15 that viscous dissipation is due to an established boundary layer between the drop and the substrate. The viscous dissipation term in Equation 15 would thus require a modification for Reynolds numbers lower than 500 to incorporate this effect.

The maximum spread factor of isopropanol is greater than that for water due to its larger Weber number. For impact of a liquid with We > 1, the spreading process is driven by the radial pressure gradient induced by the sudden velocity change at the impact location [23]. After the maximum spread factor is reached, the water splat recoils (Figure 5). The isopropanol case shows that less recoiling occurs after the maximum spread factor is attained.

## 4.2 Heat transfer

### 4.2.1 Effect of Biot number and drop properties

Figure 6 shows the influence of the Biot number on the four liquids. As the Biot number increases, heat transfer occurs more rapidly between the substrate and the drop for all cases. This can be verified by the location of typical isotherms for Bi = 1 and 100. It is interesting to notice that the temperature gradients in the solder drop are in the radial direction, while isopropanol and FC-72 splats exhibit axial temperature gradients. This is due to the higher thermal diffusivity of the solder: during the impact, the solder drop assumes a doughnut shape, with high-temperature fluid continuously supplied to the center region, so that the splat periphery is rapidly cooled by contact with the low-temperature substrate. The association of this flow pattern and the higher thermal diffusivity results in radial temperature gradients.

In later stages of spreading for isopropanol and FC-72, the heat flux across the interface is mainly governed by conduction through the substrate. The low values of thermal conductivity (Table 1) for these two liquids result in axial thermal gradients for both.

The occurrence of phase change (if any) can be predicted from Figure 4 and Figure 6. For all four liquids, when the Biot number is large (Bi = 100), the largest temperature change is seen at the periphery of the drop. This implies that the drop will begin to solidify or evaporate at its periphery first. When Bi ~ 1 the convective heat flux is reduced and the orientation of the isotherms in Figure 6 show that phase change will occur simultaneously over the *entire* contact surface between the droplet and substrate.

### 4.2.2 Effect of droplet liquid on temperature change in splat

The thermal diffusivities of the four liquids are listed in decreasing order in Table 1. Accordingly, variations of temperature inside the splat occur more rapidly for higher values of thermal diffusivities. This can be quantified analytically by considering the splat and substrate as semi-infinite bodies. The analytical solution of the transient 1-D heat conduction problem in a semi-infinite medium that is initially at a uniform temperature $T_{1,0}$ and is put in contact at time t = 0 with a semi-infinite body at another temperature $T_{2,0}$ is [27]:

$$\frac{T(z,t)-T_{1,0}}{T_{2,0}-T_{1,0}}=\mathrm{erfc}\left(\frac{z}{2\sqrt{\alpha t}}\right) \tag{16}$$

Considering the splat as a semi-infinite medium and using the analytical approach in Equation 16, we can determine analytically the axial splat thickness corresponding to a 20% temperature change at the time corresponding to the maximum extension of spreading (Table 4). For Bi = 100 this thickness is 0.96, 0.66 and 0.47 μm, for water, isopropanol and FC-72 respectively. Numerically, the thickness corresponding to a 20% change in temperature at the maximum extension of spreading can be determined from the simulations. These thickness values are 0.93, 0.64 and 0.41 μm for water, isopropanol and FC-72 respectively. In both the analytical and numerical approach, the time to reach the maximum spreading is obtained from the simulations as 24, 35 and 41 μs, for water, isopropanol and FC-72 respectively. Results for solder are not compared because thermal gradients are in the radial direction. The comparison between the analytical and numerical results for the thickness corresponding to a 20% change in temperature gives therefore reasonably consistent results (within 15% error, Table 4), provided the thermal diffusivity is not too important. In the case of solder for example, thermal diffusivity is about 1000 times higher than isopropanol and FC-72, which induces vertical isotherms: therefore no comparison is possible between the analytical model and the measurement in this case, but only comparison between the numerical calculation and the measurement.

## 5 Feasibility of experiments

A key objective of this work is to study the feasibility of using a recent laser-based temperature measurement technique [21] together with the numerical simulations. This coupled study will provide data with unprecedented temporal and spatial resolution on the behavior of interfacial heat transfer during droplet impingement on a substrate. The measurement technique is a laser-based thermoreflectance technique that measures the temperature at the fluid-substrate interface [21]. This technique is being modified to probe the temperature with an improved temporal resolution of 1 μs and a spatial resolution of 15 μm.

The setup is shown in Figure 7. A low-power He-Ne laser and a silicon photodiode are used to monitor the real-time reflectivity of the interface Both the droplet and substrate have a temperature-dependent refractive index, with the result that temperature changes in the droplet and substrate induce a reflectivity change of a laser beam incident on the droplet-substrate surface.. By measuring the change in intensity of light reflected from the interface, the temperature at the interface can be obtained. The measured temperature change $\Delta T$ is proportional to the photodiode voltage change $\Delta V$ and can be determined as follows [21]:

$$\Delta T = \frac{R_0}{V_0\left[\frac{\partial R}{\partial n_l}\frac{\partial n_l}{\partial T} + \frac{\partial R}{\partial n_s}\frac{\partial n_s}{\partial T}\right]}\Delta V \tag{17}$$

where $R$ is the reflectivity, $n$ the refractive index, and subscripts $l$ and $s$ are liquid and substrate, respectively. Since $\frac{\partial n_s}{\partial T}$ is typically much less than $\frac{\partial n_l}{\partial T}$, the variation of substrate reflectance is negligible in comparison with the variation of droplet reflectance.

Such a non-intrusive method is an ideal candidate for local and transient interface temperature measurements. Matching experimental and numerical temperatures (with the Biot number as a parameter) will allow the determination of the interfacial heat transfer coefficient.

### 5.1 Selection of liquids

The determination of the most appropriate liquids for the measurement of heat transfer coefficient can be helped by simulations showing how the temperature history of the droplet-substrate interface evolves during impingement. Such information is shown in Figure 4 and Figure 6. For example, the solder splat exhibits strong variation in the interface temperature in the radial direction for Biot numbers in the range 100. Solder is thus a strong candidate for experiments studying the spatial variation of interfacial heat transfer coefficient. On the other hand, both isopropanol and FC-72 spread more (28% to 40%) than solder (Figure 6), which implies that a proportionally larger droplet area will be available for the laser measurement. It is worth mentioning that the spreading evolution (diameter at the interface vs. time) can be measured using the same laser technique. For water, the maximum spreading of 64 μm is relatively small and the radial interface temperature is moderate, which represents a combination of both the solder and isopropanol/FC-72 behavior. As a conclusion, isopropanol can be used for testing the method, while solder and FC-72 will be tested because of their practical relevance.

## 5.2 Error induced by the spatial and temporal resolution of the measurement

This numerical study also provides estimates of the needed spatial and temporal resolution of the laser measurement to accurately capture key features of the fluid and thermal dynamics. For example, the entire spreading and cooling of a solder drop takes less than 100 μs with maximum spreading diameter of 76 μm (Figure 4). The experimental method is expected to provide an estimated temporal resolution of 1 μs and spatial resolution of 15 μm, corresponding to the circular laser spot at the droplet-substrate interface. It is worth estimating the error induced by the spatial-averaging due to the extension of the spot size, as well as the error induced by the time-averaging. This is shown in Figure 9, where spatially averaged temperature profiles (simulations of the expected outcome of the experiment) are compared with the numerically obtained temperatures. In the spatially-averaged profile, numerical temperatures are averaged within successive 15 μm spot as follows (Figure 8):

$$T_{avg,spatial} = \frac{\sum_{k=1}^{n} \left\{ T_{k,\mathrm{int}} \left( A_{k+1} - A_k \right) \right\} + A_1 T_{s1} + (\pi s^2 - A_n) T_{s2}}{\sum_{k=1}^{n} (A_{k+1} - A_k) + A_1 + \left( \pi s^2 - A_n \right)} \tag{18}$$

where $n$ is number of grid points inside the spot; $T_{k,\,\mathrm{int}}$ is the linearly interpolated temperature value at the middle of the segment joining two consecutive grid points: $T_{k,\,\mathrm{int}} = (T_k + T_{k+1})/2$; $T_{s1}$ and $T_{s2}$ are the temperatures at the intersection of the r-axis and the circular measurement spot (Figure 8). The area $A_k$ is determined by the intersection of the circular spot and the disk defined by the $k_{th}$ isotherm in the r-θ plane, located at a radial distance of $r_k$, and $s$ is the radius of measurement spot. If $e$ is distance between the spot center and the origin, $A_k$ can be expressed as follows [28]:

Case I: If $r_k > |e-s|$, then

$$A_k = r_k^2\left(\cos^{-1}\frac{e^2+r_k^2-s^2}{2er_k}\right) + s^2\left(\cos^{-1}\frac{e^2+s^2-r_k^2}{2es}\right) - \frac{1}{2}\sqrt{(-e+r_k+s)(e+r_k-s)(e-r_k+s)(e+r_k+s)} \tag{19}$$

Case II: If $r_k \le |e-s|$, then

$$A_k = \pi r_k^2 \tag{20}$$

Figure 9 compares the interface temperature obtained directly from the numerical simulation with the spatial-averaging procedure corresponding to a laser measurement (Equation 18) for all four liquids on the maximum extension of their spreading. The center location of the spatially-averaged spots $e$ is varied from 0 to $r_c$-$s$, with a resolution of 1 μm. The actual and spatially-averaged temperatures are shown as solid and dashed lines, respectively. In all cases the spatially averaged temperatures are in very good agreement with the numerical values. The numerical studies provide therefore insight the uncertainty of the experiment.

Similarly, to estimate the uncertainty induced by a measurement with a temporal resolution of $\delta t$, corresponding to the available experimental setup, numerical temperatures are averaged within $\delta t$ as follows:

$$T_{avg,temporal}(\frac{t_s + t_e}{2}) = \frac{1}{\delta t}\int_{t_s}^{t_e} T(t)dt \qquad (21)$$

where, $t_s$ and $t_e$ are the start and end time within which the temperature value is measured and $t_e - t_s = \delta t$.

Figure 10 shows the evolution of the interface temperature at $R = 0$ as a function of time for water. The solid curve shows the numerical results while the symbols simulate a temperature measurement with $\delta t = 1$, 5 and 10 μs. It appears that any of these resolutions is suitable for the temperature measurement at times larger than 10 μs after impact. However, the height of the very short transient temperature peak occurring in the first 3 μs after impact is underestimated by about 25%, even if the highest available temporal resolution available experimentally (1 μs) is used.

## 6 Conclusions

A numerical investigation of the fluid mechanics and heat transfer for a liquid microdroplet impacting on a substrate at a different temperature has been performed. In particular the effects of interfacial heat transfer, droplet spreading, and temperature variation at the interface are assessed. The liquids investigated are eutectic lead-tin solder (63Sn-37Pb), water, isopropanol and FC-72. Among the liquids, the spreading of FC-72 is the largest because of its larger Weber number. The interfacial Biot number is shown to control the location of the onset of phase change: for instance phase change is shown to happen at the droplet periphery if the Biot number is sufficiently large (Bi > 100). The numerical results are compared with published experimental results as well as an elementary analytical analysis. A key objective of this work is to assess the feasibility of a novel laser-based measurement technique to measure interfacial temperature at the droplet-substrate interface, with a high temporal and spatial resolution of respectively 1 μs and 15 μm.  To assess the feasibility of this technique, numerical results are used to predict the droplet spreading and temperature history. These numerical results are used to determine if the expected spatial and temporal limitations of the experimental technique will be sufficient to adequately resolve the transient temperatures at the droplet-substrate interface. The initial conclusions are that the experimental technique will be able to accurately capture the temperature history at the droplet-substrate interface, given the available temporal and spatial resolutions.  The results also show that the eutectic solder is the best candidate to measure radial temperature variations, while FC-72 and isopropanol exhibit larger spreading diameters and thus are natural candidates for preliminary experiments.

## 7 Acknowledgements

The authors gratefully acknowledge financial support for this work from the Chemical Transport Systems Division of the US National Science Foundation through grant 0336757.

# 9 Figure captions

Figure 1: Problem definition and initial mesh

Figure 2: Comparison of meshes generated by (a) advancing front method (b) *Mesh2D*.

Figure 3: (a) Grid independence study: Variation of z-axis contact point with time for different numbers of nodes in the splat (b) Time-step independence study: Variation of z-axis contact point with time for different time steps.

Figure 4: Spreading, recoiling and splashing of solder drop. Isotherms (on left hand side) and streamlines (on right hand side) are shown for 0.0 to 57.6 μs (Bi = 100). Splashing occurs at this latter time.

Figure 5: Evolution of spread factors with time for all four liquids.

Figure 6: Effect of Biot number on splat shape and temperature distribution for solder, water, isopropanol and FC-72. Isotherms (on left hand side) and streamlines (on right hand side) are shown at the maximum extension of spreading of corresponding splats.

Figure 7: A schematic diagram of proposed experimental set up.

Figure 8: Geometry used to calculate average temperature inside a laser measurement spot in r-θ plane.

Figure 9: Comparison of actual and spatially-averaged temperature results simulating the measurement of a laser measurement spot size of 15 μm (Bi = 100). Solid line patterns show actual results while dashed line pattern denotes spatially-averaged results.

Figure 10: Comparison of actual and time-averaged temperature results assuming a data acquisition time of 1, 5 and 10 μs, at the location $R = 0$ on interface (for water, Bi = 100). Solid line pattern shows actual result while symbols denote time-averaged results.

# 10 Tables

**Table 1: Thermophysical propeties used in the simulations**

| Droplet | Density ($kgm^{-3}$) | Thermal conductivity ($Wm^{-1}K^{-1}$) | Specific heat ($Jkg^{-1}K^{-1}$) | Viscosity (Pa-s) | Surface energy ($Jm^{-2}$) | Initial dimension-less temperature | Thermal diffusivity ($m^2 s^{-1}$) |
|---|---|---|---|---|---|---|---|
| Solder | 8218 | 25 | 238 | $2.6 X 10^{-3}$ | 0.507 | 1.0 | $1.28X10^{-5}$ |
| Water | 997 | 0.607 | 4180 | $9.8 X 10^{-4}$ | $7.3X10^{-2}$ | 0.0 | $1.46X10^{-7}$ |
| Isopropanol | 785 | 0.17 | 3094 | $2.5 X 10^{-3}$ | $2.1X10^{-2}$ | 0.0 | $7.0X10^{-8}$ |
| FC 72 | 1680 | 0.055 | 1050 | $6.4 X 10^{-4}$ | $1.0X10^{-2}$ | 0.0 | $3.1X10^{-8}$ |
| Substrate | | | | | | | |
| F2 glass | 3618 | 0.78 | 557 | - | - | 0.0 (solder); 1.0 (water, isopropanol and FC-72) | - |

**Table 2: Dimensionless numbers for drops of different liquids**

| Droplet | Re | We | Pr | Bi |
|---|---|---|---|---|
| Solder | 1254.7 | 32.4 | $2.5 \times 10^{-2}$ | 1, 10 or 100 |
| Water | 407.4 | 27.4 | 6.7 | 1, 10 or 100 |
| Isopropanol | 128.2 | 73.9 | 44.6 | 1, 10 or 100 |
| FC-72 | 1050.0 | 336.0 | 12.2 | 1, 10 or 100 |

**Table 3: Comparison of maximum spread factor with published results**

| Droplet | Maximum spread factor ($\beta_{max}$) | | |
|---|---|---|---|
| | Present work<br>Numerical | Ref. [26]<br>Experimental | Ref. [18]<br>Analytical* |
| Water | 2.41 | 2.45 | 2.44 |
| Isopropanol | 2.52 | 2.51 | 2.01 |

* contact angle $\phi$ assumed to be $90^o$ for water and isopropanol

**Table 4: Comparison of axial thickness for a 20% change in splat obtained by numerical and analytical approach**

| Droplet | Axial thickness (in μm) Numerical | Axial thickness (in μm) Analytical | Percentage error |
|---|---|---|---|
| Water | 0.93 | 0.96 | 3.2% |
| Isopropanol | 0.64 | 0.66 | 3.1% |
| FC-72 | 0.41 | 0.47 | 14.6% |

## 11 Figures

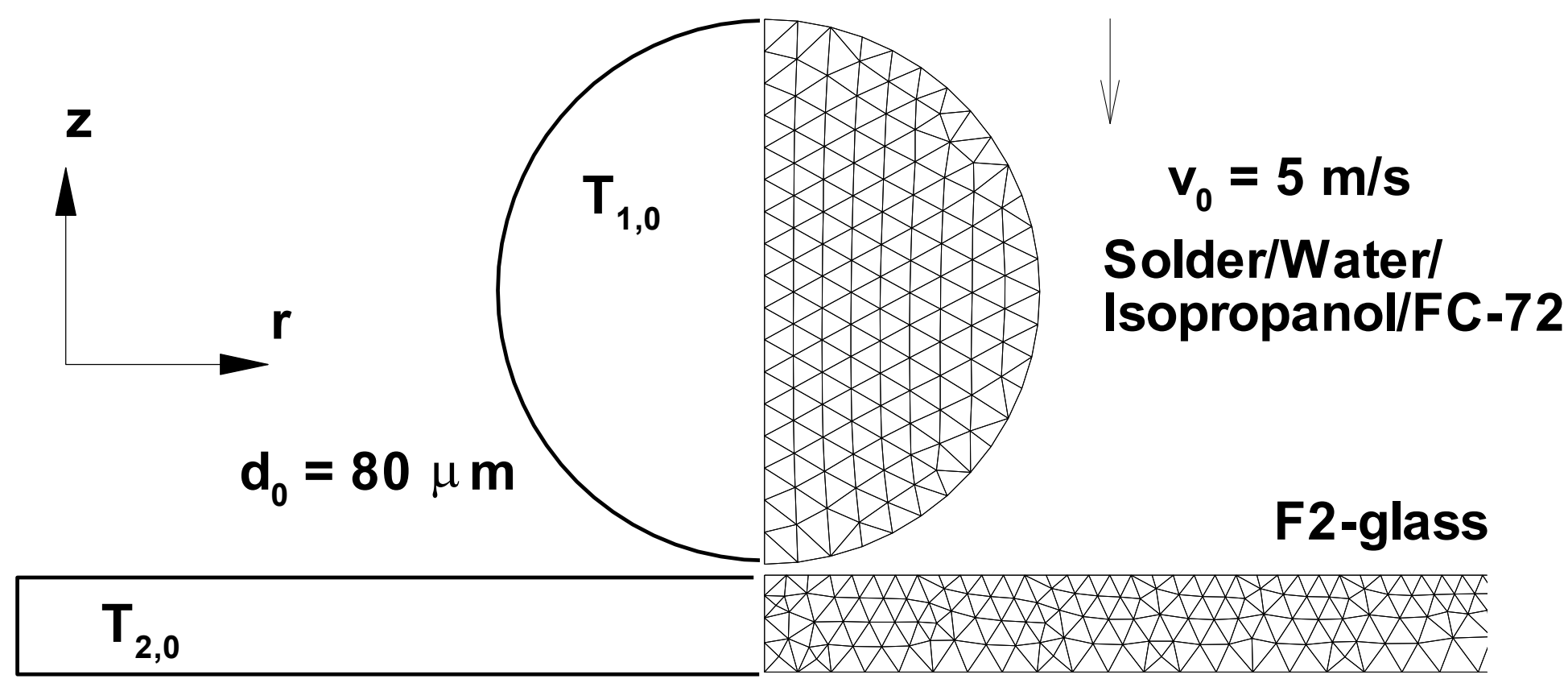

Figure 1: Problem definition and initial mesh

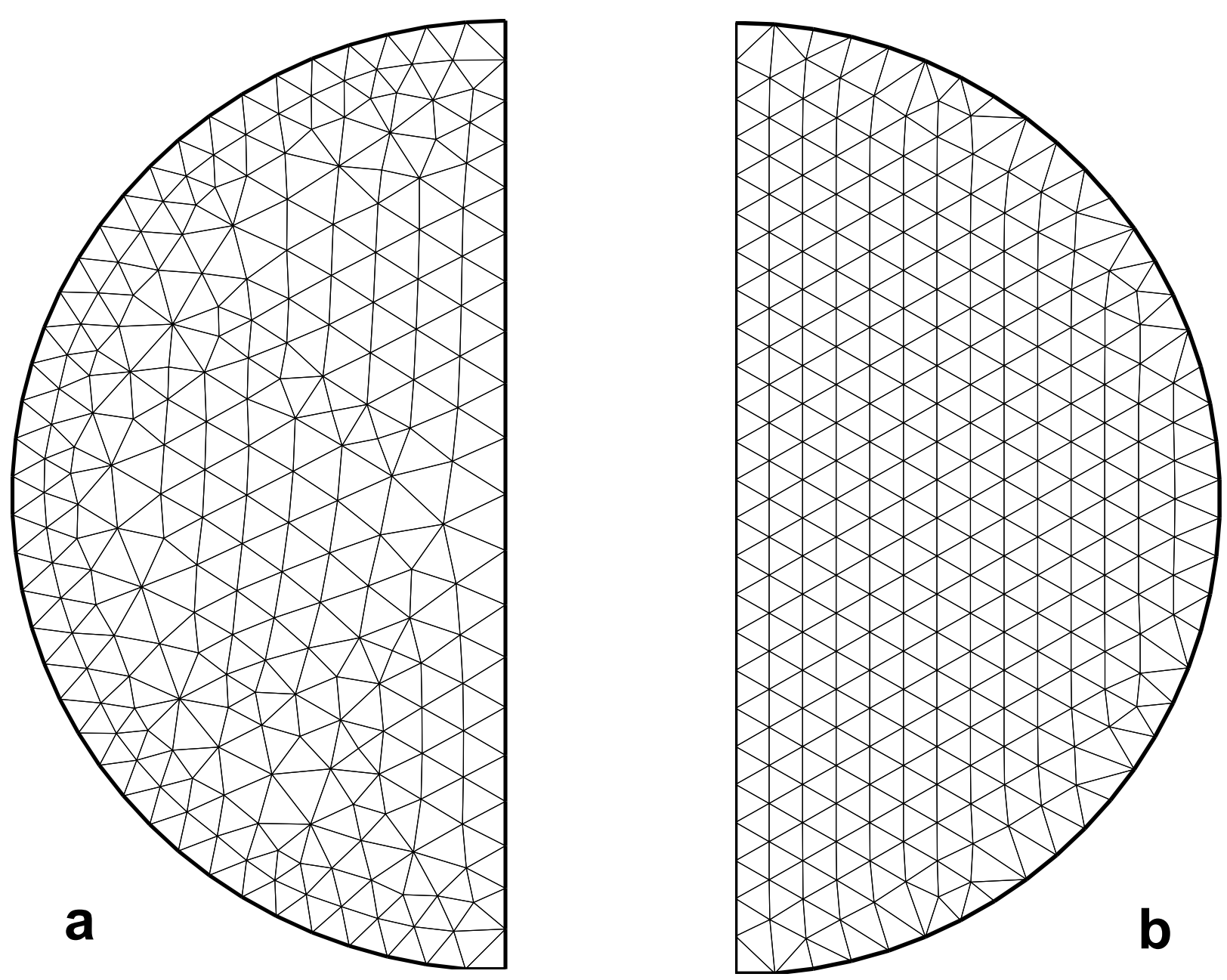

Figure 2: Comparison of meshes generated by (a) advancing front method (b) *Mesh2D*.

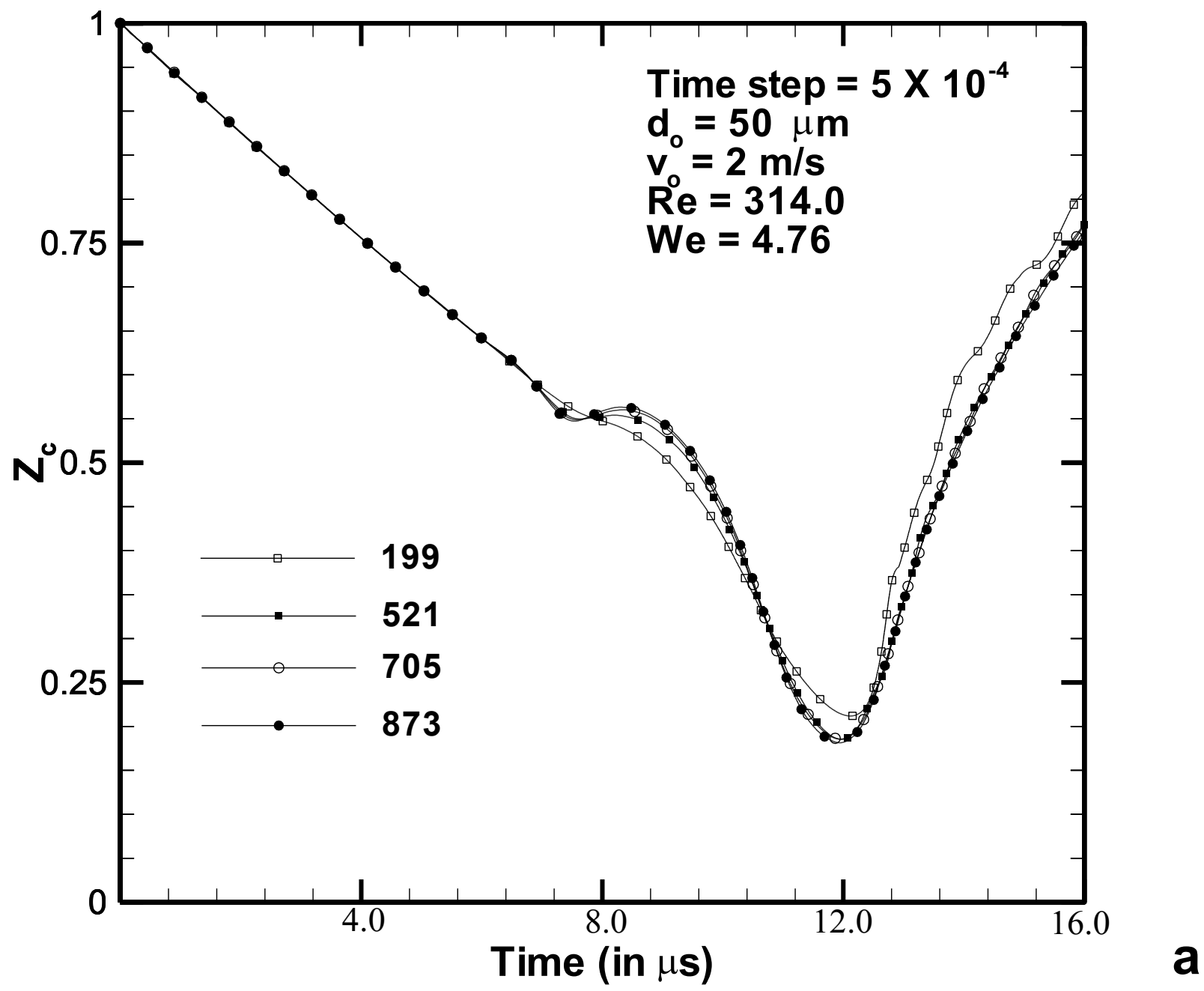

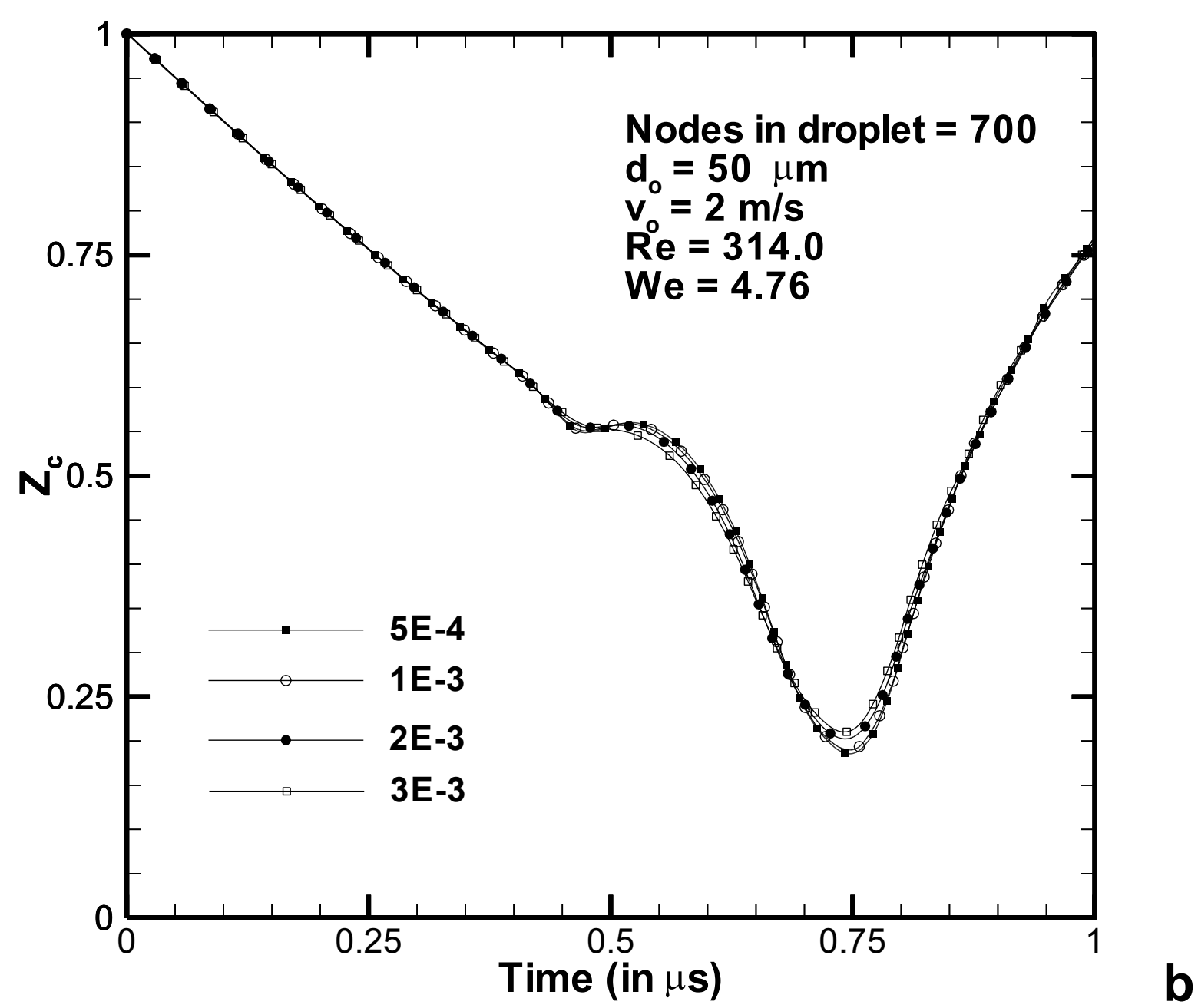

Figure 3: (a) Grid independence study: Variation of z-axis contact point with time for different numbers of nodes in the splat (b) Time-step independence study: Variation of z-axis contact point with time for different time steps.

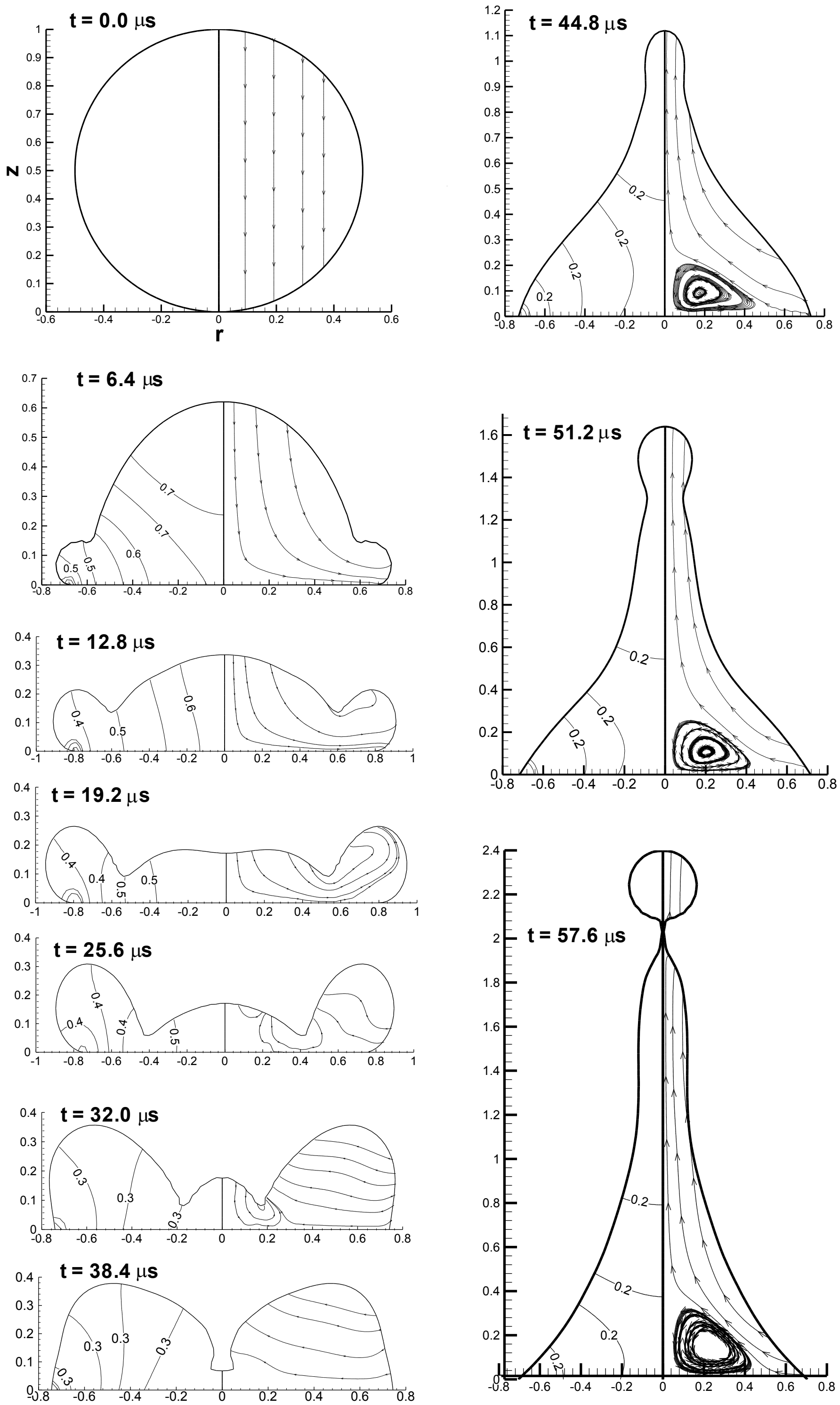

Figure 4: Spreading, recoiling and splashing of solder drop. Isotherms (on left hand side) and streamlines (on right hand side) are shown for 0.0 to 57.6 μs (Bi = 100). Splashing occurs at this latter time.

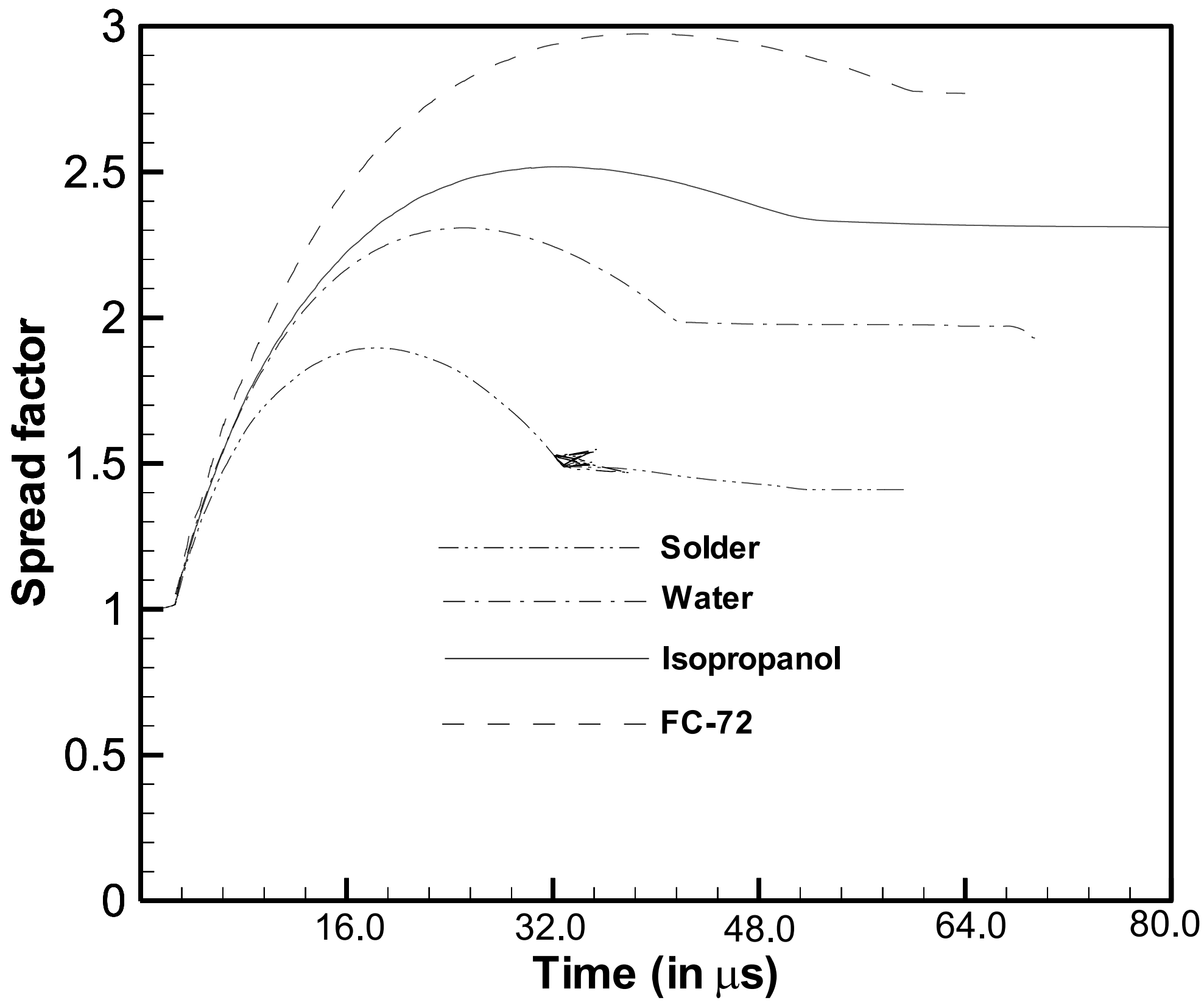

Figure 5: Evolution of spread factors with time for all four liquids.

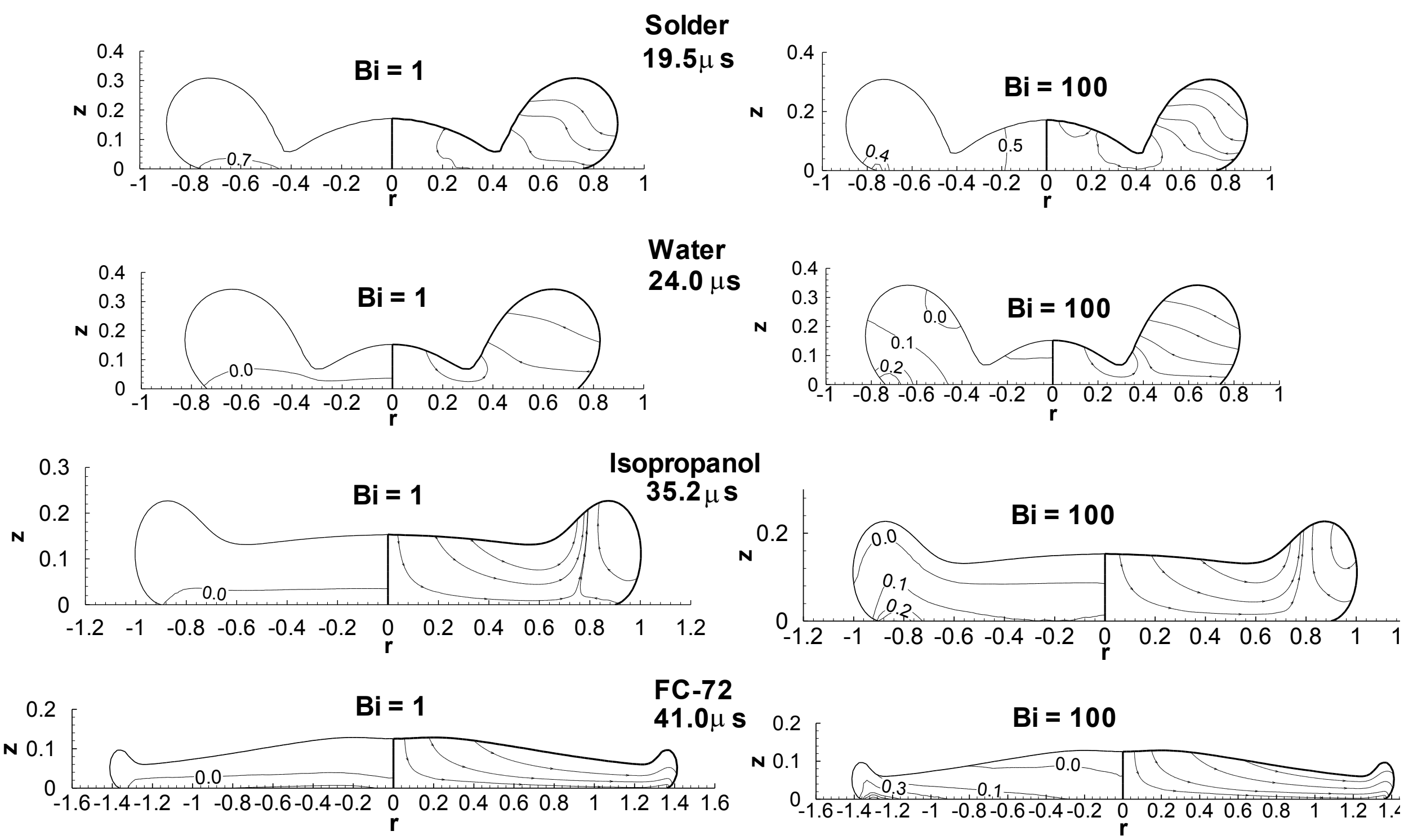

Figure 6: Effect of Biot number on splat shape and temperature distribution for solder, water, isopropanol and FC-72. Isotherms (on left hand side) and streamlines (on right hand side) are shown at the maximum extension of spreading of corresponding splats.

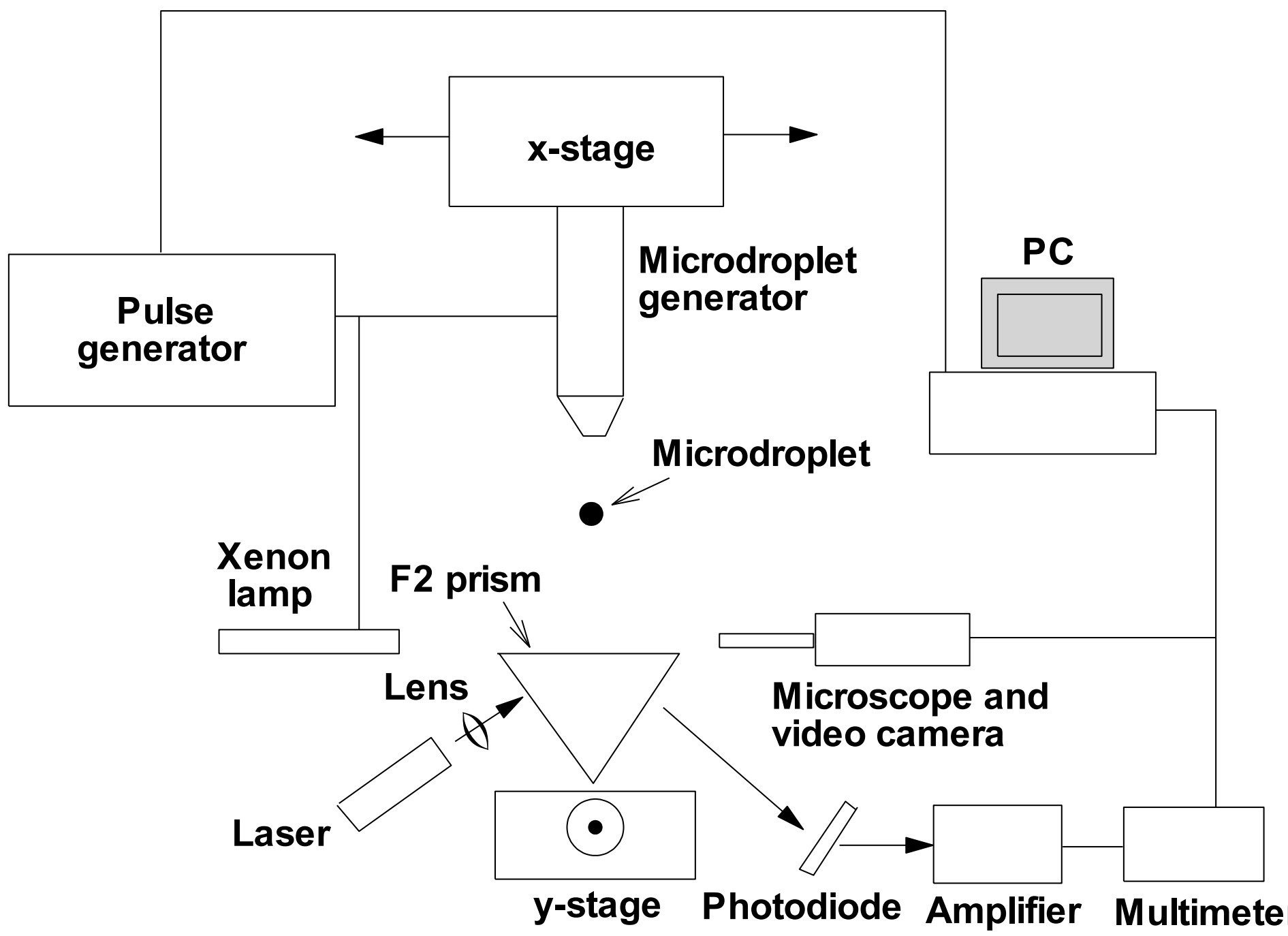

Figure 7: A schematic diagram of proposed experimental set up.

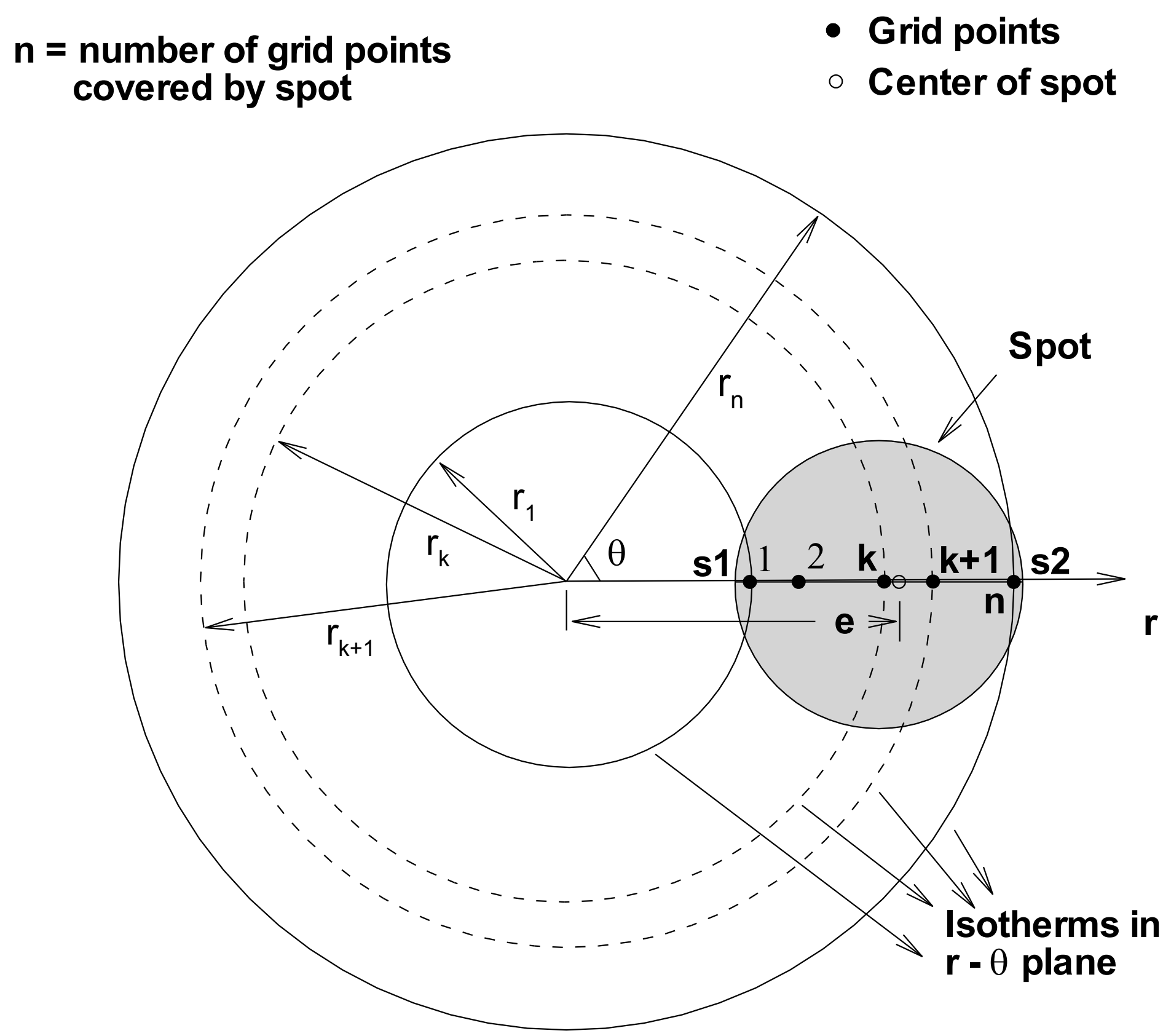

Figure 8: Geometry used to calculate average temperature inside a laser measurement spot in r-θ plane.

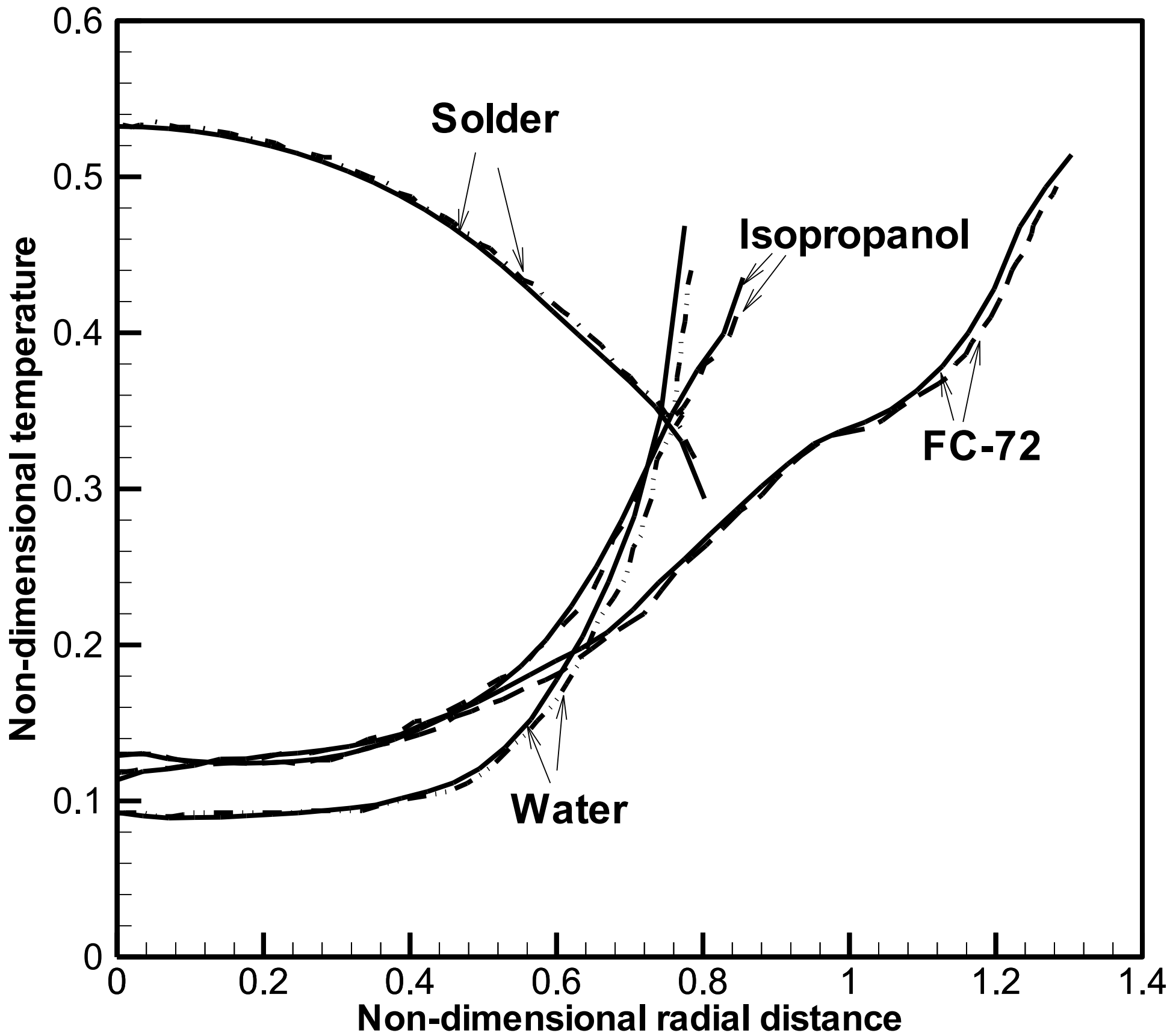

Figure 9: Comparison of actual and spatially-averaged temperature results simulating the measurement of a laser measurement spot size of 15 μm (Bi = 100). Solid line patterns show actual results while dashed line pattern denotes spatially-averaged results.

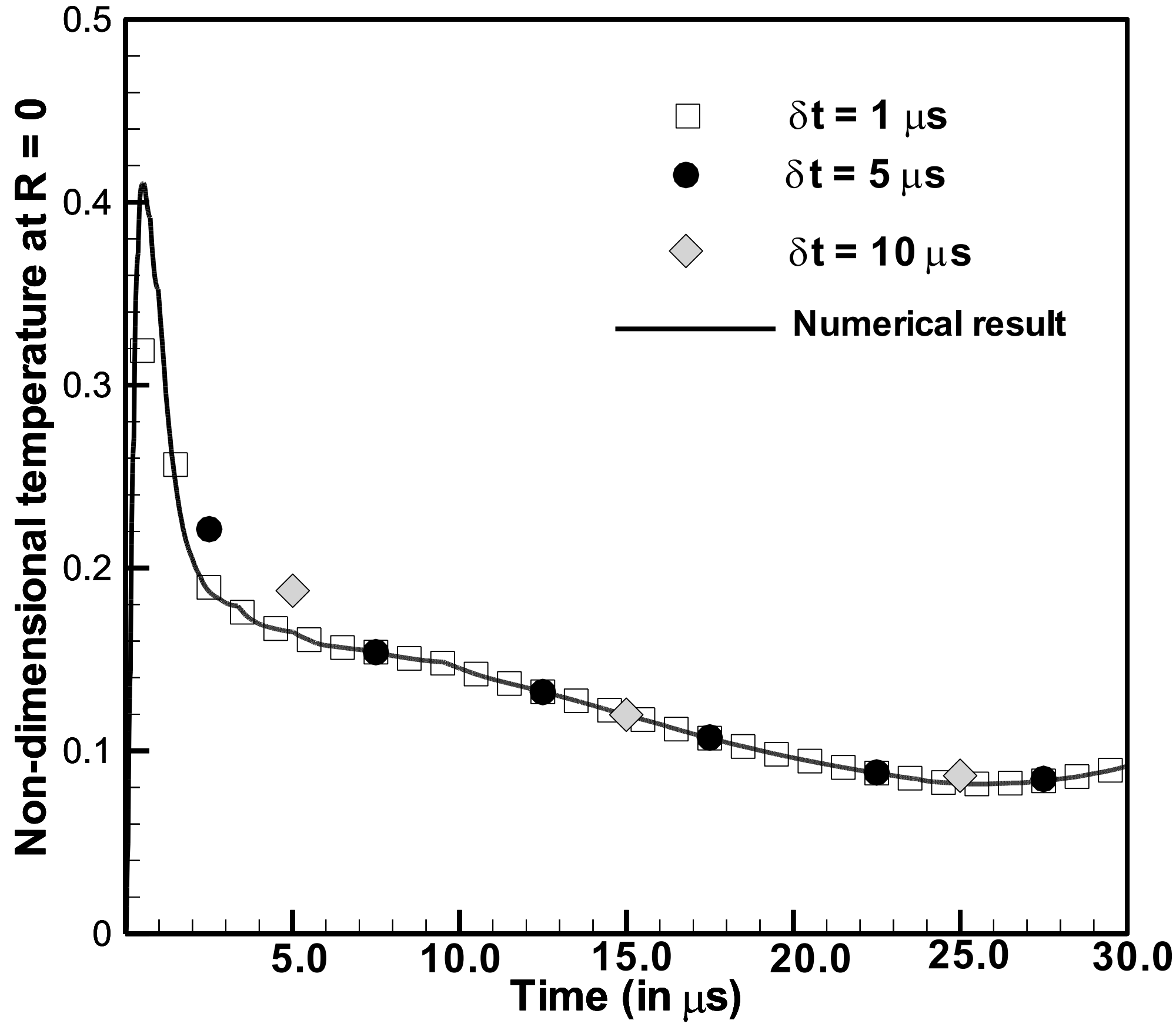

Figure 10: Comparison of actual and time-averaged temperature results assuming a data acquisition time of 1, 5 and 10 μs, at the location $R$ = 0 on interface (for water, Bi = 100). Solid line pattern shows actual result while symbols denote time-averaged results.